# The Influence of Photosynthesis on the Number of Metamers per Growth Unit in GreenLab Model

A. Mathieu *, P.-H. Cournède*, P. de Reffye**

*\* Laboratoire de Mathématiques Appliquées aux Systèmes, Ecole Centrale Paris, France*
*\*\* Laboratoire AMAP, CIRAD & INRIA, France*
*Proposed Sessions : 7 (or 6)*

## Abstract

GreenLab Model is a functional-structural plant growth model that combines both organogenesis (at each cycle, new organs are created with respect to genetic rules) and photosynthesis (organs are filled with the biomass produced by the leaves photosynthesis). Our new developments of the model concern the retroaction of photosynthesis on organogenesis. We present here the first step towards the total representation of this retroaction, where the influence of available biomass on the number of metamers in new growth units us modelled. The theory is introduced and applied to a Corner model tree. Different interesting behaviours are pointed out.

## 1. Introduction

Functional-Structural Models (FSM) aim at controlling all sides of plant growth by combining organogenesis and photosynthesis. In our proportional allocation model (see[6]), organs play their true roles as sources and sinks for biomass during the development process, and the interaction between architecture and functioning is taken into account. A good review on the FSM is found in [1]. Until now, few models exist with their corresponding simulation codes (de Reffye & al [2], Perttunen & al [3], Jallas & al [4]). Recently, de Reffye & al., see [5], proposed a FSM in the form of a dynamical system, GreenLab based on simple relevant choices from both biological and mathematical point of views. The mathematical formalism introduced allows to implement optimization and control techniques for agronomic applications.

In the previous version of the model, organogenesis was independent of photosynthesis: the number of new organs in the plant was strictly determined by genetic rules and did not vary with the production of the plant. In this paper, we present the first step towards modelling total retroaction between photosynthesis and organogenesis: the influence of available biomass on the number of metamers in new growth units is studied. For example, a beech tree in the sun can construct eleven new metamers per growth unit each year while the same tree in the shade creates two.

We start by giving a new model for the growth cycle. It allows us to introduce the influence of photosynthesis on organogenesis. Then, for a simple case, a Corner model plant, we give the mathematical equations of the plant growth and study the behaviour of the system, from theoretical and numerical point of views.

## 2. Description of the model

### 1. Plant simulation

Plants are supposed to grow in cycles and their growth is simulated with an automaton. Plants are decomposed in basic elements:

- a metamer is an internode (elementary portion of an axis) which bears lateral organs (leaves, internodes, buds, fruits);
- a growth unit is a succession of metamers set in place during a same growth cycle.

The number of growth cycles since the appearance of the organ is its chronological age. The growth unit is supposed to be preformed in the bud. Once the number of metamers is determined, plants cannot construct new ones during the cycle (no neoformation is possible).

For the sake of clarity, we suppose that the organs reach their final size as soon as they appear (immediate expansion), and they only function during one cycle. This is the case of most leafy temperate trees. Secondary growth is also neglected (no layers). However, the model can easily be extended to the general case.

## *2. Photosynthesis model*

In the photosynthesis model, the seed and the leaves are sources. Leaves, internodes, fruits and rings are sinks. Here, we do not consider the root system. From the seed, the plant constructs its first organs (following the organogenesis model). Then, during a cycle, the leaves make photosynthesis by utilizing carbon dioxide, light energy, and water. They produce the materials used for plant development. We consider a functional bud as an organ with a sink. A quantity of biomass is thus allocated to each bud, proportionally to its sink value and to the pool amount of biomass. Then, the number of organs in each bud is computed, and the matter is finally used to build them. These organs will appear in the plant at the beginning of the next cycle and so on.

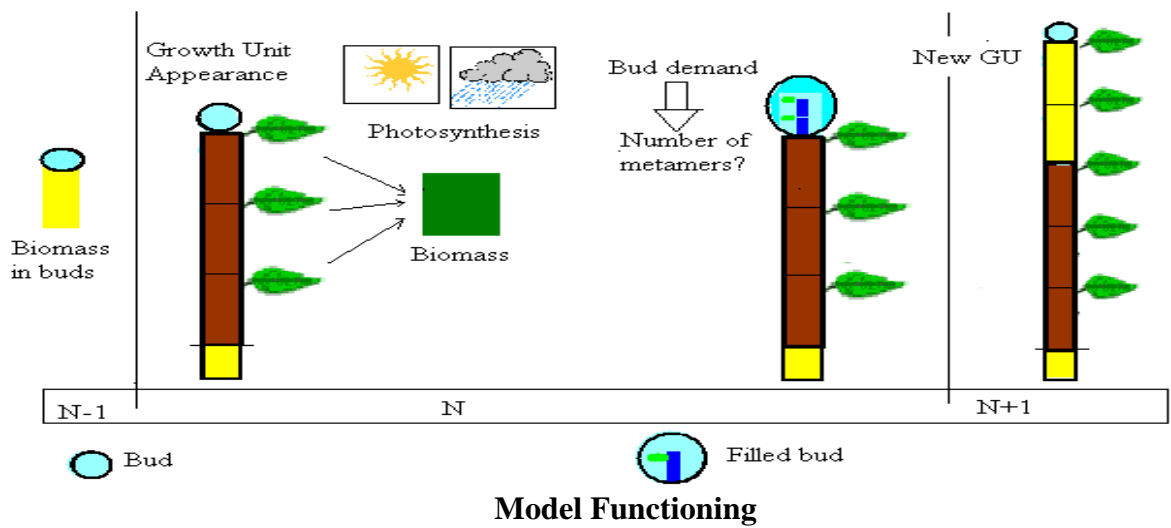

**Model Functioning**

We can sum up this functioning in several steps:
1. Calculation of the available biomass fabricated by the leaves during the cycle.
2. Calculation of the buds demand.
3. Calculation of the number of primordia in a bud, according to the ratio between biomass and demand.
4. Partitioning of the fresh matter between the future organs.
5. Breaking of the bud at the beginning of the cycle *n+1*.

Later in the development of the tree, some buds may die if there is not enough matter. Indeed, the tree will choose whether to construct less metamers or to suppress buds. This study will be presented in a next paper.

## 3. Mathematical model

We intend to equate this model of plant growth, and examine the different steps. We study a young seedling (unbranched) with only a terminal bud that is a Corner model. A single bud gives birth to a new growth unit at each cycle.

We note $N_k^{n,o}$ the number of organs $o$ of physiological age $k$ created at the beginning of cycle $n$. $p^o$ stands for the sink of an organ, ($a$ for leaf, $e$ for internode and $m$ for metamer) and $D^{b,n}$ is the bud demand at $n$. $n^a$ is the number of leaves for a metamer. $Q^n$ is the biomass produced by the leaves during cycle $n$. Finally, we introduce the resistances $r_1$ and $r_2$ for photosynthesis functions.

### 1. Biomass and Demand

Here, we suppose that only the one-cycle old leaves are active. Each leaf produces biomass that will fill the pool of reserves according to an empirical non linear function depending on its surface $S$, on hidden parameters to assess $r1, r2,$ and on an environmental parameter $E$. The empirical shape chosen here for the leaf functioning is $q^n = \dfrac{E}{\dfrac{r_1}{S^n} + r_2}$.

Eventually, the total biomass production is $Q^n = \sum_a q^n = \sum_a \dfrac{E}{\dfrac{r_1}{S^n} + r_2}$.

The demand is a scalar product between the number of organs and their sinks. Here, the matter can only be allocated to the bud, and the demand is $D^{b,n} = p_1^{BV}$. We choose now $p_1^{BV} = 1$.

### 2. Creation of metamers

According to the amount of matter allocated to the bud, we compute the number $u^{n+1}$ of new metamers in the bud. The corresponding biomass is finally used to build these metamers that appear at beginning of cycle n+ 1.

The whole available biomass allocated to the terminal bud: $q^{BV} = p_1^{BV} \dfrac{Q^n}{D^{b,n}} = Q^n$.

#### a) Determination of the number of metamers

We can compute how many metamers the bud will give birth to, with rules like: $\boxed{u^{n+1} = f(q^{BV})}$.

The function $f$ has to satisfy some constraints:
- The number of metamers in a growth unit is supposed to be bounded, the limit being a genetic datum of the plant. However, some trees may not have such limits.
- The sizes of metamers are to be taken into consideration. A tree will tend to build a supplementary metamer only when the other ones have a big enough size.

We tested functions like $f(q^{BV}) = \dfrac{1}{\dfrac{a}{(q_k^{BV})^\alpha} + b}$. Parameters are to be estimated to fit real trees.

### b) Matter distribution

Once the number of metamers is determined, the matter is divided between them. Each leaf gets $q^a = p^a \dfrac{q^{BV}}{p^m u^{n+1}}$, and we can deduce volumes and dimensions from allometry rules. We assume that the leaf thickness e is constant and the surface: $S^n = \dfrac{q^{n,a}}{e}$.

Finally, the total biomass produced by the tree is: $\boxed{Q^{n+1} = \dfrac{E n^a u^{n+1}}{\dfrac{r_1}{\dfrac{p^a}{e p^m u^{n+1}} Q^n} + r_2}}$, with $u^{n+1} = f(q^{BV})$.

## 4. Results

We studied the variation of the number of metamers $u^{n+1}$ in the growth unit of a Corner tree.

We chose $n^a = 1$ (one leaf per metamer), and note $A = \dfrac{r_1}{\dfrac{p^a}{e p^m} p^{BV}}, B = r_2$.

We can rewrite: $Q^{n+1} = \dfrac{u^{n+1} E}{\dfrac{A u^{n+1}}{Q^n} + B}$ with $u^{n+1} = f(Q^n) = round(\dfrac{(Q^n)^\alpha}{a})$.

| t | Biomass | Nb_met | Vol_met |
|---|---------|--------|---------|
| 0 | 1 | | |
| 1 | 1,21 | 2 | 0,61 |
| 2 | 1,44 | 2 | 0,72 |
| 3 | 1,68 | 2 | 0,84 |
| 4 | 2,01 | 3 | 0,67 |
| 5 | 2,37 | 3 | 0,79 |
| 6 | 2,73 | 3 | 0,91 |
| 7 | 3,08 | 3 | 1,03 |
| 8 | 3,55 | 4 | 0,89 |
| 9 | 4,02 | 4 | 1,01 |
| 10 | 4,47 | 4 | 1,12 |
| 11 | 4,87 | 4 | 1,22 |
| 12 | 5,22 | 4 | 1,31 |
| 13 | 5,76 | 5 | 1,15 |
| 14 | 6,24 | 5 | 1,25 |

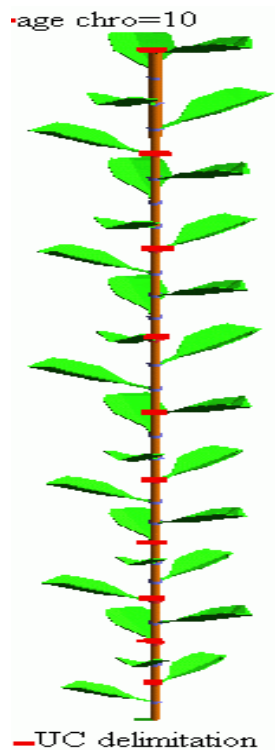

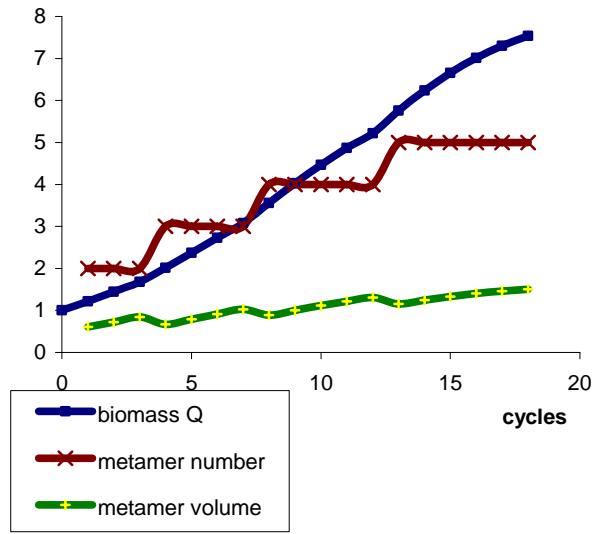

We remark the number of metamers increases, the limit being 6.

Parameters have a great influence on the plant behaviour. The more pertinent values for our study are a>0 and $1 > \alpha > 0$. In this example, we have chosen $\alpha = 1/2$, a=1/2, E=1000, A=750, B=150, $Q^0$=1; C=1/a=2; $p^a$=$p^e$=0.5; $p^{BV}$=1.

## 5. Conclusion

We have presented the bases of a theoretical model describing the influence of photosynthesis on the number of metamers per growth unit. The model can be easily generalized to more complex cases. However, it needs to be fitted to real plants, which raise a problem of choice and adjustment of its parameters. How should we define them? For example, the sinks of the buds are very important to determine the numbers of metamers. Should they be linked to the organs demand in the growth unit? Likewise, to represent a maximum of the real plants, the decision rules must be very flexible. But, the more flexible they are, the more parameters they have and the harder it is to find their values. With some parameters, we can describe a plant whose number of metamers is periodic, and some chaos may even appear! Therefore, complementary studies (implying confrontation of the model to real plants, like beech tree) are in process in order to refine the choice and definition of the model parameters.